# Sampling High Throughput Data for Anomaly Detection of Data-Base Activity

Research in Progress

**Hagit Grushka-Cohen**

Software and Information Systems Engineering, Ben-Gurion University of the Negev, Israel {hgrushka@post.bgu.ac.il}

**Oded Sofer**

IBM Security Division, Israel {odedso@il.ibm.com}

**Ofer Biller**

IBM Cyber Security Center of Excellence, Beer Sheva, Iseral {ofer.biller@il.ibm.com}

**Michael Dymshits**

Software and Information Systems Engineering, Ben-Gurion University of the Negev {m.dmshts@gmail.com}

**Lior Rokach**

Software and Information Systems Engineering, Ben-Gurion University of the Negev {liorrk @bgu.ac.il}

**Bracha Shapira**

Software and Information Systems Engineering, Ben-Gurion University of the Negev {bshapira@bgu.ac.il}

## ABSTRACT

Data leakage and theft from databases is a dangerous threat to organizations. Data Security and Data Privacy protection systems (DSDP) monitor data access and usage to identify leakage or suspicious activities that should be investigated. Because of the high velocity nature of database systems, such systems audit only a portion of the vast number of transactions that take place. Anomalies are investigated by a Security Officer (SO) in order to choose the proper response. In this paper we investigate the effect of sampling methods based on the risk the transaction poses and propose a new method for "combined sampling" for capturing a more varied sample.

# 1. INTRODUCTION

Databases lie at the heart of IT organizational infrastructure. Organizations monitor database operations in real time to prevent data leakage. Data security and data privacy protection (DSDP) systems are widely used to help implement security policies and detect attacks and data abuse. DSDP systems monitor database (DB) activity and enforce a predefined policy in order to issue alerts about policy violations and discover vulnerabilities such as weak passwords or out-of-date software. DSDP systems also apply anomaly detection algorithms in an attempt to detect data misuse, data leakage, impersonation, and attacks on database systems [1, 2].

These systems generate alerts when policy rules are violated or anomalous activities are performed [3,4,5]. Each alert demands the attention of a security officer (SO) [6] who must decide whether an alert represents a threat which should be investigated or dismissed. For investigation purposes, the security officer (SO) needs the original log data to be as informative as possible as she needs to assemble the passel in order to evaluate the whole picture. Organizations also save this data for future investigation in case a breach is discovered later. To address these needs some industrial DSDP systems maintain an archive of log data describing past transactions, and these logs are then passed on to an anomaly detection system to identify suspicious activity.

Unlike the network domain, in the data-base security domain organizations prefer not to admit the attacks hence there is no indication to which or what portion of the anomalies are "true positives". The databases monitored by DSDP systems serve thousands of users making up to a hundred thousand transactions per second. Storage is limited and can become a prohibitive cost for these systems, yet the amount of data that is logged affects the quality of anomaly detection and the ability to investigate historical data when a breach is discovered. Current solutions for the limited storage space are based on defining a policy to govern which portion of the transactions will be saved and audited.

Most processing solutions for fast streams of log data focus on compressing the data. These include various techniques of dimensionality reduction such as PCA, deep learning / auto encoding approximation methods (including sketching), and Bloom filters [7,8]. These methods allow extracting features without consuming excess memory and disk space. However, saving compressed data does not provide SOs with the information required to decide how to respond to anomalies and does not facilitate the investigation of past behavior.

Another approach to reducing the amount of data saved while preserving the full attributes is to sample transactions or activities. Techniques for sampling and their effects on anomaly detection have been studied in the domain of network traffic flow [9,10,11]. This domain is quite different from the domain of database transaction as the data is richer, containing more features, and the damage from a single transaction can be greater than the damage from a network packet.

We propose using a sampling strategy based on the perceived risk posed by each transaction to the organization. The risk can be estimated using a manually calibrated policy or estimated using a machine learning ranking algorithm such as CyberRank [12].

We seek a smart monitoring policy based on information theory sampling that will be economical (storage-wise), without compromising our ability to discover the same anomalies, and also enable us to investigate an anomaly once discovered.

The main contribution of our work is introducing new feature representing "subject matter expert knowledge" for storage reduction and improving performance of DSDP systems while maintaining the anomaly detection results and the investigative capabilities as required by the SO and the regulator. We present a sampling strategy for incorporating the risk associated with transactions \ events and preliminary results on anomaly detection using simulated data.

## 2. DATA EXPLORATION

We collected 24 hours of user activity data from a production DSDP system. The system's data is made of aggregated transactions details such as the IP address, time of the day and description of the preformed query. The users can be ether application data base user or a real user. As described in our CyberRank work [12], the user is an important entity whose behavior and activity is useful for identifying risk and controlling database transactions. When a transaction occurs, it is compared to the user's history to detect anomalies.

The data collected consists of 24 hours of monitored DB transactions made by 1,901 unique users. We identified three different user behaviors: most users have very little activity with just a few queries (less than a thousand queries per user). The second group includes users with a lot of activity and thousands of queries. User activity is described in the histogram in Figure 1.

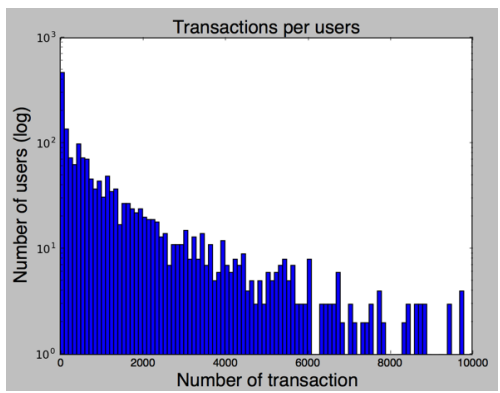

Figure 1. Histogram of user activity. Most very active users, those with over ten thousand transactions, are not depicted. The majority of the users had less than 1,000 queries in the time frame inspected

In this work we concentrate on the largest group of users which have a lower number of queries. These users are most affected from the sampling strategy as very little information is present for each user.

### 3. EXPERIMENTAL SETTING

Our objective was to compare different strategies for sampling transaction data. We look at three sampling methods:

(i) vanilla: keeping a specific portion of the transactions for sampling

(ii) risky only: sampling only from the transactions labeled as risky

(iii) combination sampling: sampling from both risky and non-risky transactions using the Gibbs sampling approach to define the portions of each transaction class in the resulting data

The DSDP system we investigated recorded the information per transaction of 20 features. Detecting anomalies with such complex features is not a trivial task, especially as it is difficult to determine

*Grushka et al./ Sampling High Throughput Data for Anomaly Detection of Data-Base Activity*what counts as a true positive anomaly without consulting the SO about each transaction (the data is unlabeled).

Since we want to investigate the impact of sampling strategy based on the new risk feature on the results of anomaly detection we concentrate on producing low-complexity data to simulate the data sampled for audit. Anomalies are introduced randomly into the data, so the anomaly detection system can be evaluated in a controlled environment (the experiment data is labeled).

### 3.1 Data creation

We represent the user as a Gaussian distribution which produces a series of observations. An anomaly is created by sampling a data point obtained from a distribution different from that of the user**.** We generated 1,000 users, and for each user a series of observations. The generated observations contained 1% anomalies, an observation generated from a different Gaussian. We added a new feature to the data - A risk label, generated randomly in an independent manner for each observation, and 30% of the observations were labeled as risky.

### 3.2 Anomaly detection

We simulated the following anomaly detection system for each user: (1) Each observation may be sampled for audit. (2) If it is sampled, it is compared to the previous series of sampled observations. (3) The user's Gaussian is estimated using the mean and standard deviation. (4) An observation is marked as an anomaly by the mock system if the observation value difference from the user's mean is greater than three standard deviations.

Each sampling method was applied three times with different random seeds, and the results were averaged. We sampled at four proportions: 35%, 30%, 25%, and 20%. The sampling posteriors (proportion of each class in the sample) for the "combination sampling" method were set at 80% for the risky class and 20% for the non-risky class

## 4. RESULTS

*Proceedings of the 11th Pre-ICIS Workshop on Information Security and Privacy, Dublin, Ireland December 10, 2016*

Without sampling and when applied to all of the data generated, the simulated anomaly detection system detected 57% of the anomalies with precision of 64%, with no significant difference between the observations labeled as high-risk or low-risk.

We compare the recall of the generated anomalies between the sampling methods, and the results are divided by class (high-risk / low-risk). The vanilla sampling method produced the same recall for both classes - as the high-risk class is less frequent, the recall for this class is low. The "risky only" sampling method achieved high recall for the high-risk class at the price of zero recall for the low-risk class. The "combination sampling" method achieves recall slightly lower than "risk only" on the high-risk class, while still capturing some of the anomalies in the low-risk class. See Figure 2 for a comparison of the recall for the two classes.

## 5. DISCUSSION

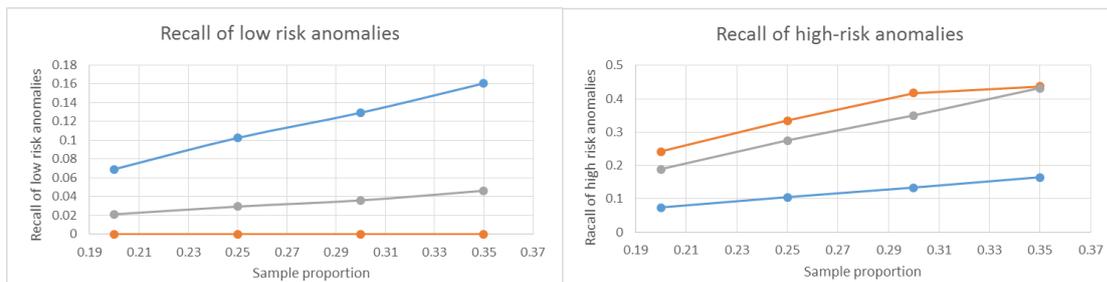

*Figure 2(a) recall for low-risk anomalies – vanilla sampling produces the best result (b) recall for high-risk anomalies – "risky only" has the higher recall.*

Choosing the right sampling strategy is an important decision when designing and implementing a DSDP system. We have shown that introducing transaction risk into the anomaly detection sampling process can significantly affect the results. In our experiment there was no underlying difference between the high and low-risk classes, we expect these transactions to behave differently in real life, amplifying the effect of the sampling algorithm on the observed anomalies.

The risk captures the likelihood that an SO would investigate the anomaly [12], however the investigation will be more thorough if low-risk transactions are also captured for the suspect user. The "risky only" method would not provide this level of resolution, while the naïve vanilla sampling would not detect most of the high-risk anomalies. Using a Gibbs sampling approach to provide

"combination sampling," guaranteeing sampling mainly from the high-risk class provides a middle ground. The proportion of each class can be tuned by the SO to fit the organizational needs (the "risky only" method is simply "combination sampling" with the proportion of high-risk samples set to one).

We are working on applying a similar type of analysis to a sample of real data, as well as more complex anomaly detection methods. Other future work should include a usage study of the effect of different sampling strategies on the adoption and use by security officers. Another venue for future work is to combine data compression methods with sampling for improving both anomaly detection and the ability to investigate afterwards.